# Generation of Vortex Beams with Strong Longitudinally Polarized Magnetic Field by Using a Metasurface

Mehdi Veysi, Caner Guclu, and Filippo Capolino*

*Department of Electrical Engineering and Computer Science, University of California, Irvine, California 92697, USA*

*Corresponding author: f.capolino@uci.edu

*Abstract-* A novel method of generation and synthesis of azimuthally E-polarized vortex beams is presented. Along the axis of propagation such beams have a strong longitudinally polarized magnetic field where ideally there is no electric field. We show how these beams can be constructed through the interference of Laguerre-Gaussian beams carrying orbital angular momentum. As an example, we present a metasurface made of double-split ring slot pairs and report a good agreement between simulated and analytical results. Both a high magnetic-to-electric-field contrast ratio and a magnetic field enhancement are achieved. We also investigate the metasurface physical constraints to convert a linearly polarized beam into an azimuthally E- polarized beam and characterize the performance of magnetic field enhancement and electric field suppression of a realistic metasurface. These findings are potentially useful for novel optical spectroscopy related to magnetic dipolar transitions and for optical manipulation of particles with spin and orbital angular momentum.

## I. Introduction

Spectroscopy systems usually work based on electric dipole transitions, which are dominant effects in interaction of molecules and atoms with electromagnetic fields. However, it would be desirable to boost the magnetic dipole transitions, which are weaker than the electric ones, to a level that can be directly detected. It is demonstrated in [1] that the ratio of magnetic dipole to electric dipole absorption rate is proportional to the square ratio of the magnetic and the electric field, $|\mathbf{H}|^2/|\mathbf{E}|^2$. Thus, detection of magnetic dipole transitions can be selectively boosted to rates comparable to electric dipole transitions by driving the particles with beams whose magnetic-to-electric-field ratio is purposely engineered. The magnetic-to-electric-field ratio is significant in the near field region of a very small circular aperture (like a fiber tip), and greatly enhances as the aperture radius decreases [2]. However, for practical aperture radii, the enhancement in magnetic field intensity is negligible [2]. For an azimuthally E-polarized beam, the magnetic-to-electric-field ratio is significantly larger than that of a plane wave $|\mathbf{H}|/|\mathbf{E}| = 1/\eta_0$ where, $\eta_0$ is the free-space wave impedance [1]. Azimuthally E-polarized beams are therefore promising for microscopy and spectroscopy methods based on detection of both magnetic and electric dipole transitions. Optical circular dichroism to study a vast amount of organic chiral molecules [3] would also benefit from enhancing magnetic fields.

Azimuthally E-polarized beams can be directly generated by coherent interference of two orthogonally polarized $TEM_{01}$ laser modes [4]. In the past few decades there has been a growing interest in novel azimuthal polarizers comprising anisotropic metallic and dielectric structures with the ability to mimic polarization manipulation capabilities of the natural birefringent media. These include interferometric techniques [5], holograms [6], liquid crystal devices [7] and spatial light modulators [8]. Space-variant dielectric gratings have been also used to convert the circularly polarized incident beams into radially or azimuthally E-polarized beams [9]. Various flat optics devices can be realized by thin metasurfaces that received considerable attention recently [10]–[12]. Optical metasurfaces comprising nanoantennas offer vast flexibility in the design of space variant polarizers by spatially tailoring the polarization state of incident beam. Recently, a set up consisting of two cascaded metasurfaces separated by a linear polarizer have been proposed to generate vector vortex beams from linearly polarized incident beam [13]. Though interesting, the structure proposed in [13] is complex, bulky and not integrated.

In this paper we examine both analytically and numerically generation of azimuthally E-polarized vortex beams through interference of Laguerre Gaussian (LG) beams and study the evolution of their electric and magnetic field distributions as they propagate in a host medium. Ideally such beams possess no electric field along the beam axis where only longitudinally polarized magnetic field is present. This characteristic is the main interest of this investigation. After showing the basic principles for generating such beams, we show how these specific beams can be generated by using metasurfaces and investigate the physical parameters they should possess. The azimuthally E-polarized vortex beam is generated from a linearly polarized incident Gaussian beam passing through a flat thin metasurface made of anisotropic nanoantennas. We also show where large magnetic-to-electric-field contrast ratio is obtained with the goal to describe where only intense magnetic field is present. Finally we show how focusing the generated azimuthally E-polarized vortex beam through a high numerical aperture (NA) lens provides strong longitudinally-polarized magnetic field in a narrow spot on the beam axis where the total electric field is negligible.

## II. ANALYTICAL MODEL

Consider an azimuthally E-polarized vortex beam, whose total electric field is expressed as

$$\mathbf{E} = E_0(\rho,z)\left(-\sin\varphi\hat{\mathbf{x}} + \cos\varphi\hat{\mathbf{y}}\right) = E_0(\rho,z)\hat{\boldsymbol{\varphi}}. \quad (1)$$

Here bold letters denote vectors and carets (^) denote unit vectors. Furthermore in the following we consider time harmonic fields with $\exp(-i\omega t)$ time dependence, which is suppressed in the following for convenience. The corresponding magnetic field is then found by $\nabla \times \mathbf{E} = -i\omega\mu\mathbf{H}$ in cylindrical coordinates, yielding

$$\mathbf{H} = \frac{i}{\omega\mu}\left[\frac{\partial E_0(\rho,z)}{\partial z}\hat{\boldsymbol{\rho}} - \left(\frac{E_0(\rho,z)}{\rho} + \frac{\partial E_0(\rho,z)}{\partial \rho}\right)\hat{\mathbf{z}}\right]. \quad (2)$$

If $E_0(\rho,z)$ has a zero of order 2 or more on the beam axis at $\rho = 0$, both the electric and magnetic fields are zero along the beam axis. However, in a special case that $E_0(\rho,z)$ has a simple zero at $\rho = 0$, the longitudinal component of magnetic field is non-zero on the beam axis where the magnitude of total electric field is zero. Note that the longitudinal component of electric field is zero everywhere in the paraxial regime as it will be explained later on in the paper. This means that there is an infinite magnetic-to-electric-field contrast ratio on the beam axis. Let us rewrite the azimuthally E-polarized field in Eq. (1) as follows

$$\begin{aligned}\mathbf{E} &= E_0(\rho,z)\hat{\boldsymbol{\varphi}} = \\ &= E_0(\rho,z)\left[-\frac{e^{i\varphi}-e^{-i\varphi}}{2i}\hat{\mathbf{x}} + \frac{e^{i\varphi}+e^{-i\varphi}}{2}\hat{\mathbf{y}}\right].\end{aligned} \quad (3)$$

In Eq. (3) we can easily recognize $e^{\pm i\varphi}$ phase dependences which represent orbital angular momentum (OAM) carrying beams with OAM numbers ±1 [14] as will be explained in the following. Thus, initial characterization of the azimuthally E-polarized beams shows that they need to possess a simple zero of radial E-field profile function $E_0(\rho,z)$ and they are a superposition of OAM carrying beams. In order for the ideal azimuthally E-polarized vortex beam in Eq. (1) to be physical, its electric and magnetic fields should satisfy the wave equation. Let us resort for a moment to a linearly polarized LG laser modes with electric field $u(\rho,\varphi,z)e^{ikz}\hat{\mathbf{x}}$, where the complex scalar function $u(\rho,\varphi,z)$ satisfies the wave equation in cylindrical coordinates that under paraxial approximation [14] reduces to

$$\left[\frac{1}{\rho}\frac{\partial}{\partial\rho}\left(\rho\frac{\partial}{\partial\rho}\right) + \frac{1}{\rho^2}\frac{\partial^2}{\partial\varphi^2} + 2ik\frac{\partial}{\partial z}\right]u(\rho,\varphi,z) = 0. \quad (4)$$

The solutions $u(\rho,\varphi,z)$ in cylindrical coordinates are LG beams that exhibit an angular phase variation as $\exp(il\varphi)$ and carry $l\hbar$ OAM per photon [14], [15]. Here, $\hbar$ is reduced Planck's constant and the parameter $l$ is called OAM number or *l*-number. These LG mode solutions are expressed as [14], [15]

$$\begin{aligned}u_{l,p} &= \sqrt{\frac{2p!}{\pi(p+|l|)!}}\frac{1}{w}\left[\frac{\rho\sqrt{2}}{w}\right]^{|l|}e^{-\frac{\rho^2}{w^2}}e^{i\frac{k\rho^2}{2R}} \times \\ &\times L_p^{(|l|)}\left(\frac{2\rho^2}{w^2}\right)e^{-i(2p+|l|+1)\tan^{-1}\left(\frac{z}{z_R}\right)}e^{il\varphi}\end{aligned} \quad (5)$$

where $u_{l,p}$ has unit norm,

$$w = w_0\sqrt{1+(z/z_R)^2}, \quad R = z\left(1+(z_R/z)^2\right), \quad (6)$$

$L_p^l(\cdot)$ is the associate Laguerre polynomial given as

$$L_p^{(|l|)}\left(\frac{2\rho^2}{w^2}\right) = \sum_{m=0}^{p}\frac{(-1)^m(p+|l|)!}{(p-m)!(|l|+m)!m!}\left(\frac{2\rho^2}{w^2}\right)^m, \quad (7)$$

$p$ indicates the radial mode number, $w_0$ is the beam waist of fundamental Gaussian mode [LG mode with $(l,p) = (0,0)$], $z_R = \pi w_0^2/\lambda$ is the Rayleigh range and $k = 2\pi/\lambda$ and $\lambda$ are the wavenumber and wavelength in the host medium, respectively. To ensure that the electric field distribution in Eq. (1) constitutes a self-standing beam with an infinite magnetic-to-electric-field-contrast ratio along the beam axis, (i) it must satisfy the paraxial wave equation, thus it must be a linear combination of LG modes with *l*-numbers ±1, (ii) the radial field profile $E_0(\rho,z)$ must have a simple zero at $\rho = 0$. The transverse E-field profiles of LG modes with *l*-numbers ±1 has the proportionality $u_{\pm 1,p} \propto \rho L_p^{(|l|)}(2\rho^2/w^2)$ as $\rho \to 0$. Note that for any $p$, the first term of associated Laguerre polynomial [$m = 0$ term of the summation in Eq. (7)] is a constant equal to the combination $\binom{p+|l|}{|l|}$, which equals $p+1$ when $l = \pm 1$. As $m$ increases in Eq. (7), higher order polynomial terms proportional to $(2\rho^2/w^2)^m$ are added. Owing to the $m = 0$ term in Eq. (7), the radial field profiles $u_{\pm 1,p}/e^{\pm i\varphi}$ of all the LG modes with $l = \pm 1$ and $p = 0, 1, 2...$ have a *simple zero* as $u_{\pm 1,p} \propto \rho$ when $\rho \to 0$. Therefore, one can use pairs of LG modes $u_{1,p}$ and $u_{-1,p}$ to construct an azimuthally E-polarized beam. By substituting the radial field distribution of LG modes with $l = \pm 1$ and any arbitrary $p$ in Eq. (3) as $u_{\pm 1,p} = E_0(\rho,z)e^{\pm i\varphi}$ with

$$E_0(\rho,z) = \frac{2\rho}{w^2\sqrt{\pi(p+1)}} e^{i\frac{k\rho^2}{2R} - \frac{\rho^2}{w^2}} \times e^{-2i(p+1)\tan^{-1}\left(\frac{z}{z_R}\right)} L_p^{(1)}\left(\frac{2\rho^2}{w^2}\right), \quad (8)$$

the electric field takes the form

$$\mathbf{E} = -\frac{u_{1,p} - u_{-1,p}}{2i}\hat{\mathbf{x}} + \frac{u_{1,p} + u_{-1,p}}{2}\hat{\mathbf{y}}$$
$$= \frac{-i\sqrt{2}}{2}\left(u_{-1,p}\,\hat{\mathbf{e}}_{rh} - u_{1,p}\,\hat{\mathbf{e}}_{lh}\right) \quad (9)$$

The right-hand and left-hand circularly polarized unit vectors are defined as $\hat{\mathbf{e}}_{rh} = (\hat{\mathbf{x}} + i\hat{\mathbf{y}})/\sqrt{2}$ and $\hat{\mathbf{e}}_{lh} = (\hat{\mathbf{x}} - i\hat{\mathbf{y}})/\sqrt{2}$. Therefore one needs two LG beams with $l = \pm 1$, with opposite sense of circular polarization, to generate the required beam. Since the linearly polarized LG beams ($u_{\pm 1,p}\,\hat{\mathbf{x}}$ and $u_{\pm 1,p}\,\hat{\mathbf{y}}$) are solutions to the paraxial wave equation, any linear combination of those, including the azimuthally E-polarized vortex beam expressed in Eq. (9), is a solution as well. The longitudinal electric field and magnetic field components of azimuthally E-polarized vortex beam can be obtained from its transverse electric field components given in Eq. (9) by using Maxwell equations under paraxial approximation (See Eqs. (19)-(23) in [16]). It is demonstrated in Appendix A that, for an azimuthally E-polarized vortex beam as in Eq. (9), the longitudinal electric field component is zero everywhere in the paraxial regime. The longitudinal magnetic field is also obtained from Eq. (2) as

$$H_z = \frac{-2i}{w^2 \omega\mu\sqrt{\pi(p+1)}} e^{i\frac{k\rho^2}{2R} - \frac{\rho^2}{w^2}} e^{-i2(p+1)\tan^{-1}\left(\frac{z}{z_R}\right)} \times$$
$$\times \left(L_p^{(1)}\left(\frac{2\rho^2}{w^2}\right)\left[2 + \rho^2\left(\frac{ik}{R} - \frac{2}{w^2}\right)\right] - \frac{4\rho^2}{w^2}S\right) \quad (10)$$

where

$$S = \begin{cases} 0 & p = 0 \\ L_{p-1}^{(2)}\left(\frac{2\rho^2}{w^2}\right) & p \geq 1 \end{cases} \quad (11)$$

Note that the strength of longitudinal magnetic field $H_z$ is inversely proportional to the square of beam waist as in Eq. (10), implying that a tightly focused beam can boost the longitudinally polarized magnetic field level. Figure 1 shows the magnitude of total electric field and longitudinal magnetic field at the transverse plane ($z = 0$) for an azimuthally E-polarized vortex beam generated using two circularly polarized LG beams with $l = \pm 1$ and $p = 0$. The electric vector field distribution is symmetric about the beam axis as shown in Fig. 1 (a). It is observed that the total electric field of azimuthally E-polarized beam vanishes on the beam axis whereas the longitudinal component of magnetic field takes its maximum value. Note that the amplitude of the total electric field is maximum at the radius $\rho = w_0/\sqrt{2}$. The normalized magnetic-to-electric-field contrast ratio $|\eta_0 \mathbf{H}|/|\mathbf{E}|$ versus radial distance from the beam axis is shown in Fig. 2. Note that the normalized magnetic-to-electric-field-contrast ratio tends to infinity on the beam axis because the electric field vanishes there.

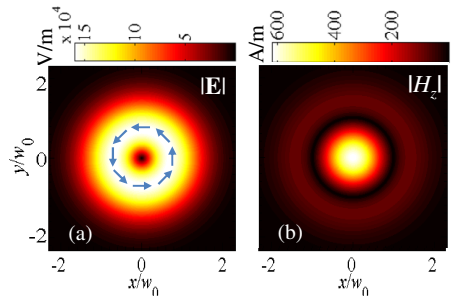

Fig. 1. (a) Magnitude and polarization of the total electric field and (b) the longitudinal magnetic field of an azimuthally E-polarized vortex beam (generated from two circularly polarized LG beams with $l = \pm 1$ and $p = 0$) at the $z = 0$ transverse plane, with $w_0 = 0.5\lambda$ at $\lambda = 6\mu m$.

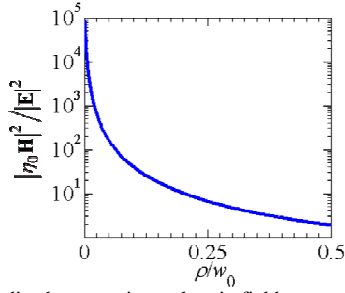

Fig. 2. Normalized magnetic-to-electric-field contrast ratio of an azimuthally E-polarized vortex beam (generated from two circularly polarized LG beams with $l = \pm 1$ and $p = 0$ ) with $w_0 = 0.5\lambda$ at $z = 0$ transverse plane as a function of radial position.

## III. METASURFACE-BASED AZIMUTHAL POLARIZER

In this section, a thin plasmonic metasurface capable of converting a linearly polarized incident beam into an azimuthally E-polarized vortex beam is conceived. In the previous section, we demonstrated that an azimuthally E-polarized vortex beam can be realized by interference of two circularly polarized LG modes with *l*-numbers of $\pm 1$. We introduce first the method for tailoring the transverse phase profile of a circularly polarized wave which is then used for generating LG modes with $l=\pm 1$. We assume that a linearly polarized incident wave, $\mathbf{E}^i$, propagating in +z direction normally illuminates a metasurface made by an arbitrary shaped nanoantenna. As shown in Fig. 3, the reference metasurface unit cell is rotated by angle $\psi$ about the propagation axis ($z/z'$-axis). We define local (primed) coordinate system attached to the unit cell and global (un-primed) coordinate system with the common $z \equiv z'$ axis. It is assumed that the transmission matrix in local (primed) coordinate system related to a unit cell with a specific orientation is known. The transmission matrix in the global coordinate system can be calculated using coordinate transformation forth and back as [17], [18]

$$\begin{pmatrix} E_x^t \\ E_y^t \end{pmatrix} = \begin{pmatrix} \cos\psi & -\sin\psi \\ \sin\psi & \cos\psi \end{pmatrix} \begin{pmatrix} T_{x'} & 0 \\ 0 & T_{y'} \end{pmatrix} \times \\ \times \begin{pmatrix} \cos\psi & \sin\psi \\ -\sin\psi & \cos\psi \end{pmatrix} \begin{pmatrix} E_x^i \\ E_y^i \end{pmatrix}. \quad (12)$$

Here the subscript *x*, and *y* denote, respectively, the *x*- and *y*-components of electric field and the superscripts *t* and *i* stand for the transmitted and incident fields, respectively. Note here that the transmission matrix of reference unit cell case is particularly chosen with null off-diagonal entries. The linearly polarized incident wave is decomposed into right (*rh*) and left (*lh*) hand circularly polarized waves as

$$\mathbf{E}^i = E_{rh}^i \hat{\mathbf{e}}_{rh} + E_{lh}^i \hat{\mathbf{e}}_{lh} . \quad (13)$$

The transmitted wave

$$\mathbf{E}^t = E_{rh}^t \hat{\mathbf{e}}_{rh} + E_{lh}^t \hat{\mathbf{e}}_{lh} \quad (14)$$

is therefore composed of

$$\begin{aligned} E_{rh}^t &= E_{rh}^i \frac{A}{2} + E_{lh}^i \frac{B}{2} e^{-2i\psi} \\ E_{lh}^t &= E_{rh}^i \frac{B}{2} e^{2i\psi} + E_{lh}^i \frac{A}{2} \end{aligned} \quad (15)$$

in which $A = T_{x'} + T_{y'}$ and $B = T_{x'} - T_{y'}$. Note that when $A = 0$, the transmitted waves are purely circularly polarized waves whose phases, upon transmission through the metasurface, are delayed or progressed by twice the rotation angle of unit cell [17]–[19]. The azimuthal phase profile on the metasurface plane is achieved by spatially tailoring the metasurface with elements having varying transmission coefficients. In doing so, the unit cell is kept stationary and the phases of circularly polarized waves are locally manipulated by rotating the nanoantennas about their own axes. This approach is accurate when the mutual coupling between nanoantennas is negligible [17]–[19]. If the metasurface elements do not change rapidly along the surface, i.e., only a small variation is imposed to adjacent elements, the local transmission properties of the metasurface can be inferred by resorting to the concept of local periodicity [10], [17], [20]. The transmission matrix of reference unit cell can be characterized in a two-dimensional infinitely periodic setup in full-wave simulations. The phase control method described here has been explored widely in the area of transmitarray and reflectarray antenna design [17]–[20]. These concepts are used here to design a metasurface which converts a linearly polarized beam to an azimuthally E-polarized beam that possesses a longitudinally polarized magnetic field.

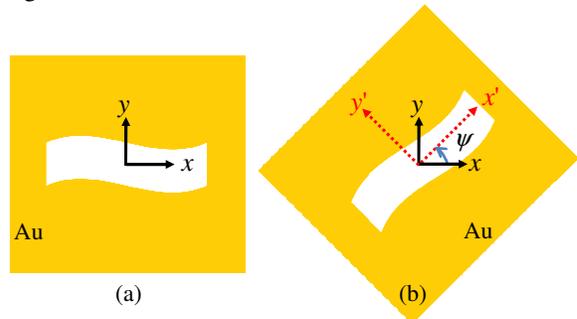

Fig. 3. Top view of an arbitrary shaped anisotropic slot nanoantenna unit cell. (a) Reference unit cell with zero rotation angle (local and global coordinate systems coincide with each other). (b) The same unit cell rotated by $\psi$ degrees, indicating both local (primed) and global (unprimed) coordinate systems.

We assume that the MS is in the *x-y* plane and it is illuminated by a normally incident Gaussian beam linearly polarized in $a_x\hat{\mathbf{x}} + a_y\hat{\mathbf{y}}$ direction, where $|a_x|^2 + |a_y|^2 = 1$ and $a_x/a_y$ is a real number. The transverse-to-*z* electric field phasor of incident wave is represented in terms of circular polarizations as

$$\mathbf{E}_{FG}^i = u_{0,0}(\rho, z)\left(a_{rh}^i \hat{\mathbf{e}}_{rh} + a_{lh}^i \hat{\mathbf{e}}_{lh}\right) \quad (16)$$

where $a_{rh}^i = (a_x - ia_y)/\sqrt{2}$ and $a_{lh}^i = (a_x + ia_y)/\sqrt{2}$ are given coefficients and $u_{0,0}(\rho, z)$ is given in Eq. (5). Using the phase control method explained above, two circularly polarized incident beams of opposite handedness are converted into circularly polarized LG modes with *l*-numbers +1 and -1 required to achieve the azimuthally E-polarized beam as shown in Eq. (9). This is achieved by setting the local rotation angle of nanoantenna element centered at the position ($\rho, \varphi$) on the surface by the angle $\psi = \varphi/2$ (thus the acquired phase becomes $e^{\pm 2i\psi} = e^{\pm i\varphi}$). This leads to spatial phase dependences of OAM states $e^{i\varphi}$ and $e^{-i\varphi}$ for left- and right-handed polarized transmitted waves, respectively. After a certain distance from the metasurface, the resultant modes are established and the total transmitted wave is the sum of four terms with regard to Eqs. (14) and (15) yielding

$$\mathbf{E}^t = \mathbf{E}_{FG}^t + \mathbf{E}_{LG}^t \quad (17)$$

where the subscripts "FG" and "LG" stand for fundamental Gaussian and Laguerre-Gaussian, respectively. The sum of fundamental Gaussian beams, whose phases are not controlled by the rotation angle as seen in Eq. (15), yields a linearly polarized beam

$$\mathbf{E}_{FG}^t = u_{0,0}(\rho, z)\frac{A}{2}\left(a_{rh}^i \hat{\mathbf{e}}_{rh} + a_{lh}^i \hat{\mathbf{e}}_{lh}\right) \\ = u_{0,0}(\rho, z)\frac{A}{2}\left(a_x\hat{\mathbf{x}} + a_y\hat{\mathbf{y}}\right) = \frac{A}{2}\mathbf{E}_{FG}^i \quad (18)$$

and the sum of LG modes with *l*-numbers +1 and -1 generated by the rotational phase control method, is represented as

$$\mathbf{E}_{LG}^t = E_1(\rho, z)\frac{B}{2}\left(a_{lh}^i e^{-i\varphi}\hat{\mathbf{e}}_{rh} + a_{rh}^i e^{i\varphi}\hat{\mathbf{e}}_{lh}\right). \quad (19)$$

Here $E_1(\rho, z) = E_{-1}(\rho, z)$ represents the radial field profile of $\mathbf{E}_{LG}^t$ transmitted beam, and it is composed of LG modes with *l*-numbers $\pm 1$, and $p = 0, 1, 2...$,

$$E_{\pm 1}(\rho, z)e^{\pm i\varphi} = \sum_{p=0}^{\infty} a_{\pm 1, p} u_{\pm 1, p} \quad (20)$$

as shown in Appendix B. For the case with fundamental Gaussian mode incidence as investigated here, the modes with $p = 0$ (with the coefficients $a_{\pm 1, 0}$) will be the dominant modes in Eq. (20). Note that $u_{+1, p}/e^{i\varphi} = u_{-1, p}/e^{-i\varphi}$, hence from Appendix B it can be shown that $a_{+1, p} = a_{-1, p}$.

Equation (17) represents the two contributions to the total transmitted wave. The first one is a linearly polarized beam whereas the second one is what we want to generate for obtaining beams with space-variant polarization. It is convenient to represent the radial and azimuthal components of the total transmitted wave $\mathbf{E}^t = E_\rho^t \hat{\boldsymbol{\rho}} + E_\varphi^t \hat{\boldsymbol{\varphi}}$ as

$$E_\rho^t = \frac{1}{2}\left[Au_{0,0}(\rho, z)\left(a_{rh}^i e^{i\varphi} + a_{lh}^i e^{-i\varphi}\right) + BE_1(\rho, z)\left(a_{lh}^i + a_{rh}^i\right)\right] \\ E_\varphi^t = \frac{i}{2}\left[Au_{0,0}(\rho, z)\left(a_{rh}^i e^{i\varphi} - a_{lh}^i e^{-i\varphi}\right) + BE_1(\rho, z)\left(a_{lh}^i - a_{rh}^i\right)\right]. \quad (21)$$

For a purely *y*-polarized incident wave (i.e., with $a_{rh}^i = -a_{lh}^i$), when $T_{y'} = -T_{x'}$ (thus $A = 0, B = 2T_x'$), it is clear from Eq. (21) that $E_\rho^t = 0$ and an azimuthally E-polarized vortex beam is obtained. This is also confirmed from Eq. (18) where the fundamental Gaussian beam contribution vanishes, and from Eq. (19) where the LG contribution takes the form of Eq. (9) describing an azimuthally E-polarized beam. On the other hand, for a *x*-polarized incident wave with $a_{rh}^i = a_{lh}^i$, when $T_{y'} = -T_{x'}$, a pure radially polarized beam is obtained.

Our goal is to show that one can achieve high magnetic-to-electric-field contrast by creating an azimuthally E-polarized beam under *y*-polarized incident wave. In practice, guaranteeing $A = 0$ is not realistic, whereas one can implement $|A| \ll |B|$ and create a *mainly* azimuthally polarized beam. For this case, due to the interference of the *y*-polarized $\mathbf{E}_{FG}^t$ and the $\varphi$-polarized $\mathbf{E}_{LG}^t$ contribution, the transverse electric field null does not appear at $\rho = 0$ and slightly shifts on the *x*-axis (where $\hat{\boldsymbol{\varphi}} = \pm\hat{\mathbf{y}}$ and $E_\rho^t = 0$). By setting $E_\varphi^t = 0$ on the $\pm x$ axis, the null transverse E-field location is found by solving

$$\pm A u_{0,0}(\rho, z) + B E_1(\rho, z) = 0. \quad (22)$$

When $|A| \ll |B|$, one can still realize extremely small E-field close to the beam axis. On the other hand, there is a strong longitudinal magnetic field close to the beam axis due to the generation of azimuthally E-polarized beam $\mathbf{E}_{LG}^t$ in Eq. (19). Note that all the LG contributions in Eq. (20) with $l$-numbers $\pm 1$, and $p = 0,1,2...$, contribute to the strong $H_z$ on the beam axis as explained in Sec. II. Thus, the strength of total $H_z$ on the beam axis ($\rho = 0$) is found with a summation as

$$H_z = \frac{2B}{w^2 \omega \mu \sqrt{\pi}} \left( a_{lh}^i - a_{rh}^i \right) \times$$
$$\times \sum_{p=0}^{\infty} a_{1,p} \sqrt{(p+1)} \, e^{-i2(p+1)\tan^{-1}\left(\frac{z}{z_R}\right)}. \quad (23)$$

## IV. DESIGN OF A METASURFACE

The proof of the proposed concept is shown in this section for a flat azimuthal polarizer metasurface of radius $20\lambda$ designed to operate at $\lambda = 6\mu m$ to convert the linearly polarized incident Gaussian beam to an azimuthally E-polarized vortex beam. The metasurface element is a double layer double split-ring slot as illustrated in Fig. 4. The array element with rotation angle $\psi = 0°$ is characterized in an infinite array setup under normal incidence using the finite element method implemented in CST Microwave Studio frequency domain solver, and its amplitude and phase of transmission coefficients are plotted in Fig. 5 versus frequency. It is observed that at the operating frequency (50THz) the transmission coefficients for $x$- and $y$-pol. waves are of equal amplitude (with the insertion loss of 4.6dB) and have a 164° phase difference. This characteristic means that $T_x \approx -T_y$, i.e. $|A| = 0.1697$ and $|B| = 1.2072$ with $|A| \ll |B|$. Hence the resultant transmitted beam will be mainly azimuthally E-polarized with weight $B$, and there will be a remnant of Gaussian beam with linear polarization in the transverse plane, with weight $A$.

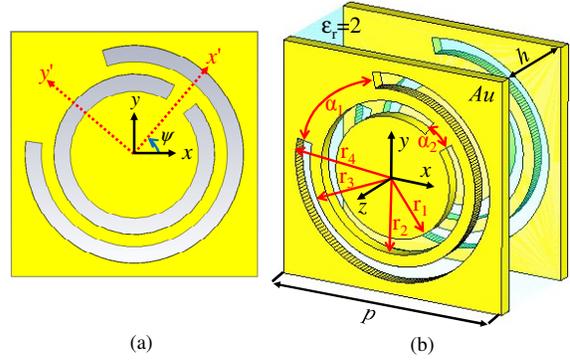

(a)          (b)

Fig. 4. (a) top view, and (b) 3D view of a double-layer double-ring slot resonator: $r_1 = 0.63\mu m$, $r_2 = 0.79\mu m$, $r_3 = 0.9\mu m$, $r_4 = 1.06\mu m$ $\alpha_1 = 66°$, $\alpha_2 = 20°$, $h = 1\mu m$, $p = 2.4\mu m$.

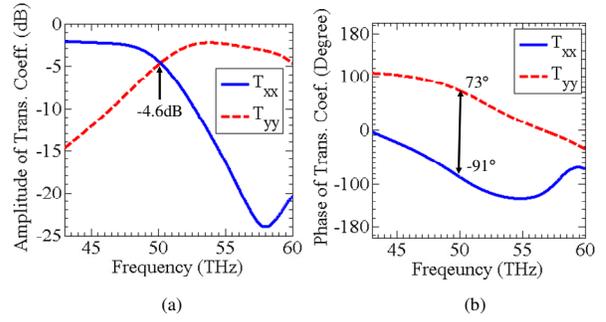

(a)          (b)

Fig. 5. (a) amplitude and (b) phase of the transmission coefficients for the $x$- and $y$-polarized incident waves. The results obtained for the reference nanoantenna with rotation angle $\psi = 0°$.

In general, other unit cell elements can be used, considering also more metasurface layers, to further decrease the insertion losses at mid-infrared and to further minimize the ratio $|A/B|$. Based on the phase control method described in Sec. III, the transmission characteristics of rotated elements are derived from the characteristics of reference one with $\psi = 0°$. One can estimate the transmitted field at every cell location by resorting to the concept of rotation and local periodicity discussed in Sec. III, without the need of characterizing all possible unit cell configurations and the whole metasurface in full-wave simulation environment. The transmitted field through the metasurface polarizer is approximated as a step function over its surface.

In the following, we utilized 2-D forward and inverse Fourier transform implemented numerically in order to model the propagation of transmitted beam through the metasurface. The transmitted electric field over a unit cell at a very short distance of the metasurface is approximated with uniform distribution over each unit cell and evaluated as

$$\mathbf{E}_{cell}(\rho_{cell}, \varphi_{cell}) = \underline{\mathbf{T}}(\psi = \varphi_{cell}/2) \mathbf{E}_g^i(\rho_{cell}, \varphi_{cell})$$

where $\underline{\mathbf{T}}(\psi)$ is the transmission tensor of the nanoantenna element rotated by the angle $\psi$, ($\rho_{cell}$, $\varphi_{cell}$) is the position of the center of unit cell. This is the field profile over the metasurface that generates the beam as in Eq. (17) further away from the surface. Propagation of the transmitted field through the metasurface is determined by first Fourier transforming the transverse piece-wise approximated electric field just over the transmission side of the polarizing metasurface assumed to be at $z = z_0$. This spectral transverse E-field is evaluated based on the formula

$$\mathbf{E}(k_x, k_y, z_0) = \int_{-\infty}^{+\infty}\int_{-\infty}^{+\infty} \mathbf{E}(x, y, z_0) e^{-ik_x x - ik_y y} dx dy . \quad (24)$$

Then the field is reconstructed by the inverse Fourier transform at any arbitrary transverse plane as

$$\mathbf{E}(x, y, z) = \frac{1}{4\pi^2} \times \\ \times \int_{-\infty}^{+\infty}\int_{-\infty}^{+\infty} \left[\mathbf{E}(k_x, k_y, z_0) e^{ik_z(z-z_0)}\right] e^{ik_x x + ik_y y} dk_x dk_y \quad (25)$$

where

$$k_z = \begin{cases} \sqrt{k_0^2 - (k_x^2 + k_y^2)} & k_0^2 \geq (k_x^2 + k_y^2) \\ i\sqrt{(k_x^2 + k_y^2) - k_0^2} & k_0^2 < (k_x^2 + k_y^2) \end{cases} \quad (26)$$

The double integrals in Eqs. (24) and (25) are efficiently calculated by using a 2D FFT algorithm, where the size of entire spatial domain is $81.92\lambda \times 81.92\lambda$ with spatial resolution of $\lambda/50$. The metasurface located at $z = z_0$ covers a circular area with a diameter of $40\lambda$ and the transverse E-field is set to zero outside of the metasurface area. Note that the evanescent near-field components are significant at distances very close to metasurface. The incident wave is a linearly polarized Gaussian beam and its electric field and power density at the beam center are $1\,\text{V/m}$ and $1.3\,\text{mW/m}^2$, respectively. The total incident beam power is $3\times10^{-11}\,\text{W}$. The beam waist of the incident Gaussian wave is set equal to the radius of azimuthal polarizer metasurface ($20\lambda$) so that 86% of the total incident beam power illuminates the metasurface. Via numerical implementations of the plane wave spectrum computations [Eqs. (24) and (25)], we first show the generated azimuthally E-polarized beam's field intensities and its evolution, and then the same beam is focused using a lens, to further boost the magnetic-to-electric-field contrast.

In Fig. 6 (a) and (b), the intensities of total electric field and longitudinal magnetic field (polarized along the propagation axis) of the generated azimuthally polarized beam at three transverse planes $z = 0.5\lambda$, $5\lambda$, and $10\lambda$ are reported, respectively. The beam clearly has an electric field null and a hot spot of the longitudinal magnetic field at the center on all transverse planes. These features are broadened in space as the azimuthally polarized beam propagates and diverges. For demonstration purposes, the transverse electric field direction at a constant radius is superimposed in the plots in Fig. 6 (a), which clearly represents an azimuthally E-polarized vortex beam.

Next, we investigate the features of azimuthally E-polarized vortex beam when a focusing lens of radius $20\lambda$ is placed $0.5\lambda$ away from the polarizer surface. In order to focus the azimuthally E-polarized beam, a hyperboloid phase profile is added to the spatial field distribution on the lens plane [21]. The intensity of focused beam is then found in any transverse plane by numerically implementing the plane wave spectrum computations as in Eqs. (24) and (25). In Fig. 7, the magnitude of the total electric field (left), the longitudinal magnetic field (middle), and the normalized ratio of the total magnetic field to total electric field (right), of the focused azimuthally polarized beam at the lens focal plane are reported using lenses with different NAs.

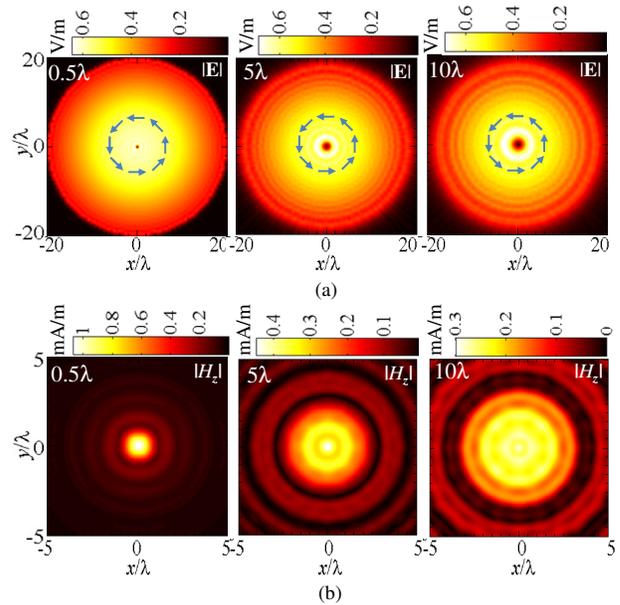

Fig. 6. (a) Intensity of the total electric field and (b) the longitudinal magnetic field of the generated azimuthally E-polarized vortex beam at different transverse planes: 0.5λ, 5λ, and 10λ away from the polarizer metasurface. The blue arrows indicate the orientation of transverse electric field vector at each plane.

The resultant tightly focused azimuthally E-polarized vortex beam creates a strong longitudinally polarized magnetic field in a very narrow spot (vortex region) where the magnitude of total electric field is negligibly small. Note that the electric field null slightly shifts away from the exact origin, this is due to the "leakage" of the original linearly polarized Gaussian beam in Eq. (18) through the metasurface as we will discuss in details regarding Fig. 9. Focusing the azimuthally E-polarized vortex beam through a lens with either NA of 0.45 or 0.7 enhances the magnetic field levels up to 7.2 to 18 times, respectively. Moreover this results in boosted levels of $|\eta_0\mathbf{H}|/|\mathbf{E}|$. Even though we observe circular regions of $|\eta_0\mathbf{H}|/|\mathbf{E}|$ maxima in Fig. 7 the magnetic field is maximum at the central region. In both cases the null of electric field and the maximum of longitudinal magnetic field are close to the center, because of the interference of the fundamental Gaussian beam with the LG modes.

In Fig. 8, the total electric field intensity, the longitudinal magnetic field intensity and the magnetic-to-electric-field-contrast ratio on the *x-z* plane are reported, clearly showing the large magnetic field and ratio $|\eta_0\mathbf{H}|/|\mathbf{E}|$ values on the beam axis along the depth of focus. The null of electric field along the *z*-axis region is clearly visible, where the magnetic field reaches maximum. In Fig. 9, we report the normalized magnetic-to-electric-field-contrast ratio $|\eta_0\mathbf{H}|/|\mathbf{E}|$ versus the radial position on the *x*-axis for the same cases in Fig. 7. The normalized contrast ratios exhibit maxima of 1600 and 1684 for lenses with NAs of 0.45 and 0.7, respectively. The location of maxima moves toward to the origin with increasing NA. The dislocation of maxima from the origin in the *x-y* plane is attributed to the linearly *y*-polarized Gaussian wave leaking through the metasurface due to $A \ne 0$ in Eq. (18). Indeed, the azimuthally E-polarized beam is also *y*-polarized along the *x*-axis, which explains why the two fundamental Gaussian and the LG beams interfere there. Moreover, the leaked linearly polarized beam also possesses a very small longitudinally polarized electric field component at $\rho \ne 0$, therefore the normalized magnetic-to-electric-field-contrast ratio assumes large but finite value at maxima, and does not go to infinity on the axis as it would occur for a symmetric and ideal azimuthally polarized LG beam.

Note that both the azimuthal polarizer and lens cannot be simply integrated into a single metasurface by using the double-layer double split-ring element, with the phase control method explained in Sec. III. This method dictates that the phase shift introduced to right- and left-hand circularly polarized waves possess opposite signs [Eq. (15)]. Therefore, if lensing was integrated in the metasurface, the phase shift distribution designed for converging the right-hand circularly polarized wave would lead to a divergent left-hand circularly polarized beam and vice versa due to the opposite sign of phases introduced in Eq. (15). Thus the azimuthally E-polarized beam could not be synthesized since both right- and left-hand circularly polarized beams should have the same amplitude distribution over any transverse plane [see Eq. (9)].

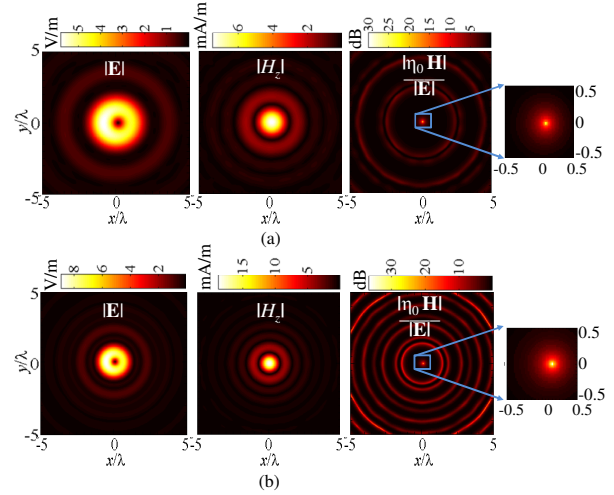

Fig. 7. The total electric field |**E**| (left), the longitudinal magnetic field |$H_z$| (middle), and the normalized magnetic-to-electric-field ratio $|\eta_0\mathbf{H}|/|\mathbf{E}|$ (right) of a tightly focused azimuthally polarized beam at the focal plane for lenses of radius $20\lambda$ with different NAs: (a) NA=0.45 ($f=40\lambda$), and (b) NA=0.7 ($f=20\lambda$). The magnetic-to-electric-field contrast ratio is normalized to its value for plane wave.

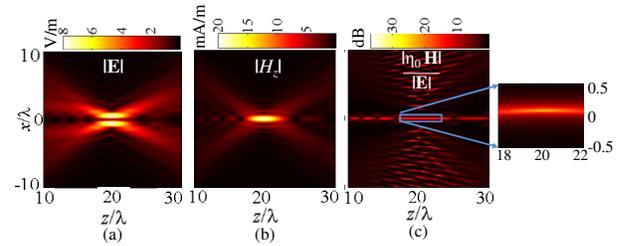

Fig. 8. Simulation results for fields in the *longitudinal* plane generated by a system composed by a polarizer metasurface followed by a lens. The polarizer and lens are placed at $z=0$, and $z=0.5\lambda$, respectively. The focal length of lens with NA =0.7 is $f=20\lambda$. (a) Magnitude of the total electric field and (b) the longitudinally polarized magnetic field, of tightly focused azimuthally E-polarized beam in the longitudinal *x-z* plane. (c) Normalized magnetic-to-electric-field ratio, showing large values along the z-axis region.

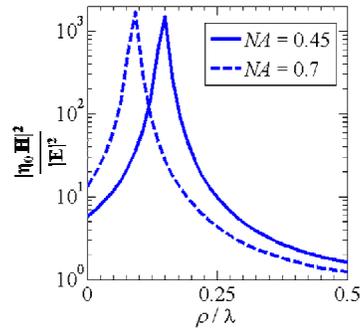

Fig. 9. Normalized magnetic-to-electric-field contrast ratio of a tightly focused azimuthally E-polarized beam as in Fig. 7, for the two lenses considered. The field is evaluated on the focal plane of each case, as a function of radial coordinate, for $\varphi = 0°$.

## V. CONCLUSION

We have investigated how to form an electromagnetic beam with a very large magnetic-to-electric-field ratio. It is demonstrated, both theoretically and numerically, that azimuthally E-polarized vortex beams acquire a strong longitudinal magnetic field on the beam axis where the electric field vanishes. We have shown the properties a metasurface should possess in order to form such a beam and explored a specific metasurface implementation made of an array of double split ring slots. Focusing the resultant azimuthally E-polarized vortex beam significantly boosts the magnetic-to-electric-field-contrast ratio in a narrow spot on the propagation axis. The performance is limited by the fundamental Gaussian beam suppression as the beam traverses through the metasurface. Such beams may find interesting applications in the optical manipulation of particles with optical magnetic polarizability. It may also open the way to future spectroscopy systems based on magnetic dipole transitions.

ACKNOWLEDGMENT

The authors would like to thank the Computer Simulation Technology (CST) for providing CST Microwave Studio.

APPENDIX A: VANISHING LONGITUDINAL COMPONENT OF ELECTRIC FIELD FOR AN AZIMUTHALLY E-POLARIZED VORTEX BEAM

We derive here the longitudinal electric field component of an azimuthally E-polarized vortex beam. The transverse electric field of an azimuthally E-polarized beam given in Eq. (9) is rewritten as

$$\mathbf{E}^t = E_x \hat{\mathbf{x}} + E_y \hat{\mathbf{y}} \quad (A.1)$$

where $x$ and $y$ components of electric field are

$$E_x = \frac{i}{2}(u_{1,p} - u_{-1,p}), E_y = \frac{1}{2}(u_{1,p} + u_{-1,p}) \quad (A.2)$$

By defining $u' = u_{1,p}/e^{i\varphi} = u_{-1,p}/e^{-i\varphi}$, the azimuthally E-polarized electric field in Eq. (A.1) can be rewritten as

$$\mathbf{E} = u'(\rho,z)(-\sin\varphi\hat{\mathbf{x}} + \cos\varphi\hat{\mathbf{y}}). \quad (A.3)$$

The longitudinal electric field component can be then found from the transverse electric field components using Maxwell equations under paraxial approximations as [16]

$$E_z = \frac{i}{k}\left(\frac{\partial E_x}{\partial x} + \frac{\partial E_y}{\partial y}\right) =$$
$$= \frac{i}{k}\left(-\frac{\partial(u'(\rho,z)\sin\varphi)}{\partial x} + \frac{\partial(u'(\rho,z)\cos\varphi)}{\partial y}\right). \quad (A.4)$$

After taking the derivatives and using the chain rule $\frac{\partial u'}{\partial x} = \frac{\partial u'}{\partial \rho}\frac{\partial \rho}{\partial x}$, the Eq. (A.4) is further simplified into

$$E_z = \frac{i}{k}u'(\rho,z)\underbrace{\left(\frac{\partial\cos\varphi}{\partial y} - \frac{\partial\sin\varphi}{\partial x}\right)}_{=0} +$$
$$+ \frac{i}{k}\frac{\partial u'(\rho,z)}{\partial\rho}\underbrace{\left(\cos\varphi\frac{\partial\rho}{\partial y} - \sin\varphi\frac{\partial\rho}{\partial x}\right)}_{=0} = 0 \quad (A.5)$$

which demonstrates that the longitudinal component of electric field is zero everywhere in paraxial regime. Note that this salient feature of azimuthally E-polarized beams is valid in the paraxial regime regardless of the radial mode number $p$ of constituent LG modes with $l = \pm 1$.

APPENDIX B: PROJECTION OF TRANSMITTED FIELD ONTO LG MODES

We show how to calculate the radial field profile of higher order LG modes generated by manipulating the phase distribution of the incident fundamental Gaussian mode $u_{0,0}$ [given in Eq. (5) for $(l,p)=(0,0)$] upon transmission through a proper surface. When an azimuthal phase profile $e^{il_1\varphi}$ is added upon phase manipulation through a surface, one has a total field equal to $u_{0,0} e^{il_1\varphi}$ which itself does not constitute an individual LG mode solution of the paraxial wave equation. For example, LG modes of order $l_1$ are characterized by a phase distribution of $e^{il_1\varphi}$, however the field radial profile of LG modes of order $l_1$, $u_{l_1,p}$, differ greatly from that of the incident fundamental Gaussian mode $u_{0,0}$. The field profile $u_{0,0} e^{il_1\varphi}$, on the other hand, generates all the LG modes $u_{l_1,p}$ with $p = 0,1,2,...$ . Therefore, the total field profile of generated beam is represented as

$$u_{0,0}e^{il_1\varphi} = \sum_{p=0}^{\infty} a_{l_1,p} u_{l_1,p} \quad (B.1)$$

which is a weighted summation of those LG modes $u_{l_1,p}$, with mode coefficients $a_{l_1,p}$. The mode excitation coefficients $a_{l_1,p}$ are found by taking the projection of transmitted field profile $u_{0,0} e^{il_1\varphi}$ onto the LG modes $u_{l_1,p}$ owing to the orthonormality of LG modes [22]

$$a_{l_1,p} = \int_0^{2\pi}\int_0^{\infty} u_{0,0} e^{il_1\varphi} u_{l_1,p}^* \rho\,d\rho\,d\varphi. \quad (B.2)$$

Note that due to the orthogonality of LG modes, no LG mode with $l \neq l_1$ can be generated by a phase profile $e^{il_1\varphi}$. As used in Eq. (19), the established beam's amplitude distribution $E_{l_1}(\rho,z)$ can be defined using

$$E_{l_1}(\rho,z)e^{il_1\varphi} = \sum_{p=0}^{\infty} a_{l_1,p} u_{l_1,p}. \quad (B.3)$$

In general, the coefficients $a_{l_1,p}$ depend strongly on the initial field profile impinging on the azimuthal polarizer metasurface (here taken as $u_{0,0}$), moreover they can be calculated numerically for any arbitrary incident field profile. Note that reversing the azimuthal phase profile added to the original field profile, which in turn becomes $u_{0,0}\,e^{-il_1\varphi}$, would result in the same radial field profile $E_{-l_1}(\rho,z) = E_{l_1}(\rho,z)$ which can be easily concluded using the identity $u_{l_1,p}/e^{il_1\varphi} = u_{-l_1,p}/e^{-il_1\varphi}$ in Eqs. (B2) and (B3).

In a more general setting, the azimuthal polarizer metasurface also scales the field strength as it manipulates the phase profile of transmitted beam. These coefficients are provided for the specific case in Eq. (19).